\documentclass[12pt,a4paper]{article}
\usepackage{psfig}

\renewcommand\thesection{\arabic{section}.}
\renewcommand\thesubsection{\thesection\arabic{subsection}.}
\renewcommand\thesubsubsection{\thesubsection\arabic{subsubsection}.}
\renewcommand\section[1]{\vspace{\topsep}\vspace{\partopsep}
\refstepcounter{section}
{\par  \noindent\normalsize\bfseries \thesection
\hspace{1em}#1\vspace{\topsep}\par\noindent}}

\newenvironment{refs}
{\vspace{\topsep}\vspace{\partopsep}
{\par \noindent\normalsize\bfseries  References
\vspace{-\topsep}\par\noindent}
\setlength{\parindent}{-5mm}
\begin{list}{}{\topsep 0pt \partopsep 0pt \itemsep 0pt \leftmargin 5mm
\parsep 0pt \itemindent -5mm}}
{\end{list}}

\renewcommand\subsection[1]{
\refstepcounter{subsection}
{\par \protect\vspace{\topsep}\vspace{\partopsep}
 \noindent\normalsize\bfseries \slshape \thesubsection
\hspace{1em}#1\par \noindent}}

\renewcommand\subsubsection[1]{
\refstepcounter{subsubsection}
{\par \protect \vspace{\topsep}\vspace{\partopsep}
\noindent\normalsize \slshape \thesubsubsection
\hspace{1em}#1\par \noindent}}

\newfont{\sansb}{cmssbx10}
\newfont{\sans}{cmss10}

\newcommand{\vol}[2]{$\;\,$\bf #1\rm , #2.} 
\def\aap{{Astron. Astr.}}                         
\def\aapl{{Astron. Astr. (Lett.)}}                
\def\asr{{Adv. Space Res.}}                       
\def\ssr{{Space Sci. Rev.}}                       
\def\apj{{Astrophys. J.}}                         
\def\apjl{{Astrophys. J. (Lett.)}}                
\def\apjs{{Astrophys. J. Supp.}}                  
\def\apss{{Astrophys. Space Sci.}}                
\def\mnras{{MNRAS}}                               
\def\nat{{Nature}}                                
\def\teq#1{$\, #1\,$}                           
{\catcode`\@=11                                                  
\gdef\SchlangeUnter#1#2{\lower2pt\vbox{\baselineskip 0pt\lineskip0pt    
\ialign{$\m@th#1\hfil##\hfil$\crcr#2\crcr\sim\crcr}}}}           
\def\gtrsim{\mathrel{\mathpalette\SchlangeUnter>}}               
\def\lesssim{\mathrel{\mathpalette\SchlangeUnter<}}    

\def\erg{\varepsilon}

\def\taupp{\tau_{\gamma\gamma}}

\begin{document}


\begin{center}
{\large \bf Gamma-Ray Bursts Above 1 GeV\vspace{18pt}\\}
{Matthew G. Baring$^{1,2}$\vspace{12pt}\\}
{\sl $^1$LHEA, NASA/Goddard Space Flight Center, Greenbelt, MD 20770, USA\\
     $^2$Compton Fellow, Universities Space Research Association\\}
\end{center}

\begin{abstract}
One of the principal results obtained by the Compton Gamma Ray
Observatory relating to the study of gamma-ray bursts was the detection
by the EGRET instrument of energetic ($>$100 MeV) photons from a
handful of bright bursts.  The most extreme of these was the single 18
GeV photon from the GRB940217 source.  Given EGRET's sensitivity and
limited field of view, the detection rate implies that such high energy
emission may be ubiquitous in bursts.  Hence expectations that bursts
emit out to at least TeV energies are quite realistic, and the
associated target-of-opportunity activity of the TeV gamma-ray
community is well-founded.  This review summarizes the observations and
a handful of theoretical models for generating GeV--TeV emission in
bursts sources, outlining possible ways that future positive detections
could discriminate between different scenarios.  The power of
observations in the GeV--TeV range to distinguish between spectral
structure intrinsic to bursts and that due to the intervening medium
between source and observer is also discussed.
\end{abstract}

\setlength{\parindent}{1cm}

\section{Introduction}
Gamma-ray bursts (GRBs) have intrigued observers and confounded
theorists ever since their discovery over twenty years ago (Klebesadel,
Strong and Olsen 1973).  Despite a plethora of source observations by
over a dozen different experiments, these fascinating transients remain
enigmatic phenomena.  While the results obtained by the Compton Gamma
Ray Observatory (CGRO) dramatically improved our observational
database, they perpetuated the confused picture we have of GRBs, and
ushered in a new age of controversy.  Bursts emit predominantly in the
10 keV--10 MeV band (for spectra, see Band et al. 1993, or the BATSE 1B
spectroscopy catalogue of Schaefer et al. 1994), with durations
normally between milliseconds and several minutes and sometimes
millisecond variability: see the BATSE 1B catalogue of Fishman et al.
(1994) for an illustration of the complexity and diversity of their
time histories.

The contribution of CGRO to the GRB paradigm has been threefold.
First, the accumulation of a more substantial burst population by one
experiment (BATSE) has permitted a more precise determination of their
celestial angular and \teq{\log N}-\teq{\log S} distributions, leading
to the confirmation that bursts are indeed isotropic, but that they are
relatively scarce (Meegan et al. 1992) at low fluxes (i.e. presumably
large distances).  The simplest explanation for this fact is via the
hypothesis that bursts are cosmological in origin: this spawned the
subsequent tergiversation of the GRB community's perspective from the
pre-CGRO view that bursts arose near neutron stars in the disk of our
galaxy.  The second definitive advance to our knowledge of bursts by
CGRO was the detection by EGRET of hard gamma-rays well above 10 MeV
(discussed at more length in the next section) from a high percentage
of bright BATSE sources, suggesting that the GRB phenomenon is not
exclusively the domain of low energy gamma-ray astronomy.  Third, the
evidence for hard X-ray absorption features in the early KONUS
observations (see Mazets et al. 1981, and Fenimore et al. 1988 for
higher spectral resolution GINGA detections of such features) was not
reproduced in the BATSE data, substantially diminishing the strongest
evidence in the pre-CGRO era for the galactic neutron star scenario.

Despite these advances, the distance scale for bursts was still not
unequivocally determined, and GRBs remained the only class of transient
astronomical sources that were believed to emit purely in gamma-rays.
Searches in the radio, optical and X-ray bands for convincing transient
or quiescent steady-state counterparts to bursts (e.g. see papers in
Paciesas and Fishman 1991) had proved negative, and the Holy Grail of
GRB astronomy remained elusive.  This year the situation changed with
the improvement of burst localizations by BeppoSAX to a few
arcminutes.  This led to the sensational discovery of X-ray (Costa et
al. 1997) and optical (Groot et al. 1997; van Paradijs et al. 1997;
Guarnieri et al. 1997; Djorgovski et al. 1997) counterparts to the GRB
970228 event (the {\it yymmdd} notation denotes burst dates).  These
sources were ascertained to be persistent but variable, exhibiting
decays on timescales of the order of a month or so.  Since then,
another burst provided counterpart detections, GRB 970228, which
revealed Mg and Fe absorption lines in optical observations by the Keck
telescope.  The z=0.77 and 0.83 redshifts of these lines (Metzger et
al. 1997) are attributed to interstellar media associated with galaxies
intervening between the burst and the observer, or perhaps also the
host galaxy for GRB970228.  If these associated sources are indeed true
counterparts, this result provides compelling evidence for the
cosmological origin of bursts.

Given these developments, it is natural to ask what role can TeV
gamma-ray astronomy play in resolving outstanding issues and refining
our understanding of burst sources?  This paper outlines the expected
impact that positive detections of bursts by ground-based Atmospheric
\v{C}erenkov Telescopes (ACTs) can make on the gamma-ray burst field in
the foreseeable future, in particular how they can constrain model
parameters and source distances.  The EGRET observations that have
already motivated GRB searches by extant experiments are reviewed, and
the handful of models that address super-GeV emission in limited
fashions are discussed.  Spectral signatures expected from pair
production attenuation internal to sources will be summarized, and
compared with the spectral absorption anticipated from interactions
with cosmological background radiation fields.  While the recent
observations of optical counterparts to BeppoSAX bursts with inferred
significant redshifts favours an emphasis on cosmological scenarios,
bursts of galactic halo origin will also be discussed for
completeness.

\section{EGRET Observations and TeV Upper Limits}
Prior to the launch of the Compton Gamma-Ray Observatory in 1991, there
were no detections of super-GeV photons coming from gamma-ray bursts,
with the highest energy photon recorded by the GRS spectrometer on the
Solar Maximum Mission (SMM) being at around 80 MeV (Share et al. 1986);
there were numerous detections at lower energies (e.g. Nolan et al.
1983).  This prompted the popular perception that bursts were
short-lived hard X-ray/soft gamma-ray phenomena, with little relevance
to ground-based experiments.  This misconception could, in principal,
be supported by results from the BATSE instrument on CGRO, which, while
it routinely observes GRB spectra extending up to and above 1 MeV,
demonstrated that most bursts exhibit spectral steepening at a variety
of energies between 50 keV and few hundred keV (Band et al. 1993; see
also Mallozzi et al. 1995 for the ``\teq{\nu F_{\nu}}-peak''
distribution and Schaefer et al. 1994 for the BATSE 1B spectroscopy
catalogue).

The observations by the EGRET and COMPTEL experiments on board CGRO
changed this picture dramatically.  Detections of 11 bursts (e.g. see
Schneid et al.  1996 for a recent listing) by the EGRET spark chamber
and/or TASC extended the spectral range of interest to the hard
gamma-ray domain.  In addition, COMPTEL has seen (e.g. Hanlon et al.
1994) over 20 bursts in the 300 keV--15 MeV range.  EGRET has detected
emission above 50 MeV from four of the brighter GRBs triggered by
BATSE; all are consistent with power-law spectra extending to as high
as 1.2 GeV, in the case of GRB 930131 (Sommer et al. 1994), and 3.4 GeV
for GRB 940217 (Hurley et al. 1994).  The GRB 940217 source is best
known for exhibiting delayed or prolonged high energy emission,
detected 80--100 minutes (i.e. more than one full earth orbit of CGRO)
after the initial trigger, including a photon of energy 18 GeV (Hurley
et al. 1994) that is not markedly inconsistent with the extrapolation
of the power-law continuum.  Not only did this observation further
expand the spectral range of bursts, it opened up the possibility of
extended emission in the time domain.  This property has motivated
recent searches for high energy gamma-ray emission by ground-based ACTs
following notification via the BACODINE alert network (discussed by
Cline et al.  in these proceedings) of BATSE burst triggers.  Some
evidence for delayed high energy emission pre-dated GRB 940217, with
the observation (Dingus et al. 1994) of a single 10 GeV photon that
could have been associated with GRB 910503.

\begin{figure}[htb]
\vspace{0.3cm}
\centerline{\psfig{file=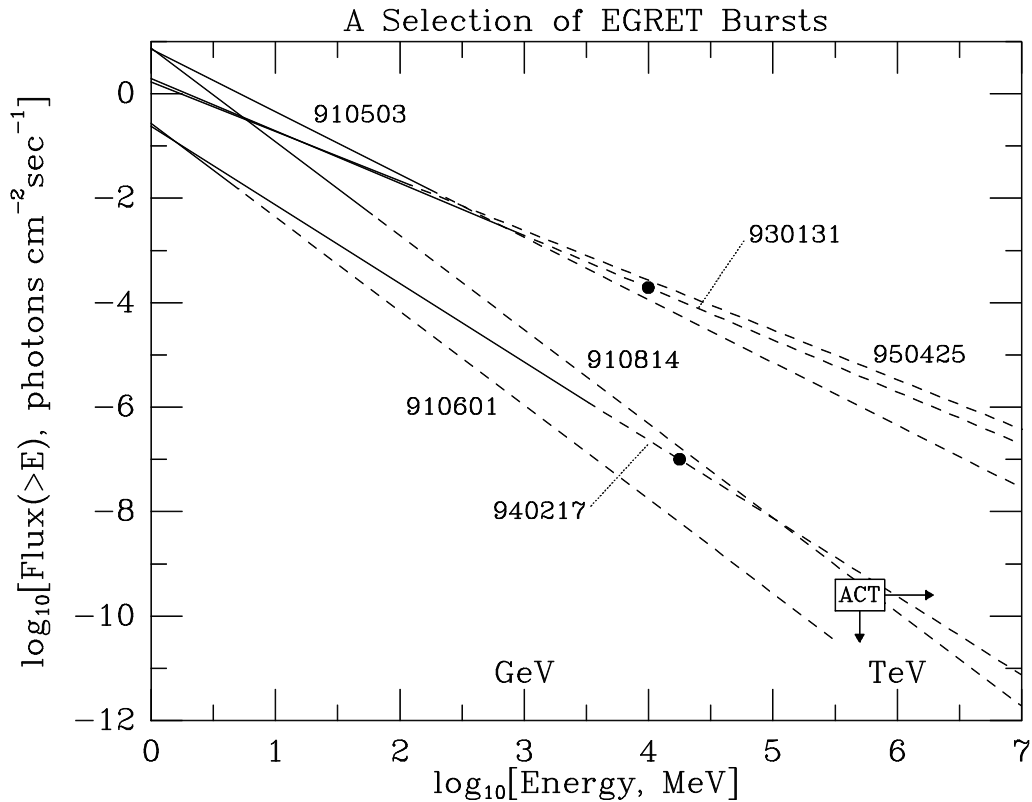,height=10.5cm}}
\vspace{-0.3cm}
\caption{The integral fluxes for six of the eleven EGRET burst
detections (a depiction after Hurley 1996).  The solid lines are the
confirmed power-laws, as taken from the data compilation in Baring and
Harding (1997b), while the dashed extrapolations are provided to
indicate fluxes in the TeV range.  The filled circles denote the single
``delayed'' photons observed to be consistent with the positions of two
bursts, GRB910503 (Dingus et al. 1994) and GRB940217 (Hurley et al.
1994).  The current threshold and sensitivity for ACT observations of
bursts is indicated by the ``ACT'' box, and is taken from the Whipple
rapid searches (Connaughton et al. 1997).
}
\end{figure}

The general relationship of EGRET source spectra to searches by
ground-based ACTs can by illustrated via the collections of hard
gamma-ray portions of GRB spectra in Figure~1.  These EGRET sources are
the ones with better detection statistics, and their source parameters
(differential fluxes at 1 MeV, spectral indices and maximum energies,
tabulated in Baring and Harding 1997b) are obtained from the source
papers of Schneid et al. (1992: GRB 910503), Kwok et al. (1993: GRBs
910601 and 910814), Sommer et al. (1994: GRB 930131, the ``Superbowl
burst''), Hurley et al. (1994: GRB 940217) and Catelli et al. (1996:
GRB 950425).  Note that a conservative approach is adopted here, with
the delayed high-energy emission being distinguished from the power-law
emission; such a choice is motivated by the possibility that the
``delayed'' photons may actually be from prolonged components that are
distinct from the lower energy power-laws.  Other EGRET sources have
statistics that are too poor to effect reliable extensions of their
spectra to the TeV range; extrapolations of the upper end of BATSE
spectra suffer from the same problem.  It is clear from the EGRET (and
also BATSE and COMPTEL) data that there have been no attenuation-type
turnovers or cutoffs observed in a GRB spectrum.  High energy gamma-ray
emission may therefore be common in bursts, if not universal: the EGRET
detection rate is consistent (Dingus 1995, though this inference is
subject to poor statistics) with all bursts emitting above about 30
MeV.

The extrapolations to TeV energies indicate that half of these bursts
would have been visible to \v{C}erenkov telescopes, assuming that these
sources intrinsically emit at such energies and that there is no
internal or external absorption of their radiation.  The fact that
these were several orders of magnitude above current ACT sensitivities
is strong motivation for a TeV search program.  Such a quest began in
earnest in early to mid 1995 by the Whipple team, made possible by the
establishment of the BACODINE notification network.  Unfortunately, this
post-dated all but one of the bursts depicted in Figure~1, so that
target-of-opportunity exposures for these sources were not possible.
Since the beginning of this program, the Whipple team has reported
(Connaughton et al. 1995, 1997; Boyle et al. 1997) only negative
results from their searches following fast telescope slewing prompted
by BATSE triggers.  The upper bounds to integral fluxes so obtained for
several bursts are approximately represented by the ``ACT'' box in
Figure~1, and pertain to observations made, at earliest, starting two
minutes after the BATSE events.  HEGRA has a similar program, also with
negative results (Padilla et al. 1997) in both tracking and archival
searches, and the status of the BIGRAT/CANGAROO initiative in this
direction is reported in Dazeley et al. (1997).  While the ACT
sensitivity box is tailored to Whipple's conditions, it provides a
rough guide to other experiments such as HEGRA and CANGAROO, which have
higher thresholds and consequently better flux sensitivity (denoted by
the two arrows attached to the box).  MILAGRO will have a slightly
worse sensitivity, but this will be compensated for by its greater
potential detection rate due to its large field of view.

\section{A Pocketful of Models}
Theoretical predictions of gamma-ray emission above 1 GeV are extremely
sparse, largely due to the paucity of data in this range.  Since the
relevant observations postdate the launch of CGRO, only models of the
cosmological {\it fireball} genre address super-GeV emission, albeit
cursorily, with limited spectral development in most cases.  Such
models, first proposed by Cavallo and Rees (1978) and considered by
numerous researchers since, usually involve a central catastrophic
event such as the gravitational coalescence of neutron star or neutron
star/black hole binaries (e.g.  Paczy\'nski, 1986; Eichler et al.
1989), or perhaps partial failure and collapse of a supernova onto its
compact core (Woosley, 1993).  In either case, roughly a solar mass of
energy is released in a very small, optically thick volume, thereby
rapidly thermalizing to relativistic temperatures (10 MeV or so,
Paczy\'nski, 1986).  This energy naturally must disperse adiabatically,
and the resulting expansion of baryons and pairs generates relativistic
bulk motions, i.e. a fireball.  Temperatures around \teq{\sim 20} keV
are usually achieved (e.g. Paczy\'nski, 1986; Goodman, 1986) in the
adiabatic cooling of fireballs, blue-shifted to MeV energies by the
bulk motion of the plasma outflow.  This is unlike the observed
non-thermal spectra of bursts, leading immediately to a problem with
pure fireball models, though radiative transfer effects (Carrigan and
Katz, 1992) may produce steep high energy tails.

In subsequent refinements to fireball models it was observed that those
with significant baryonic content are relatively inefficient at
$\gamma$-ray production (Shemi and Piran, 1990).  This property
motivated the formulation of ``blast-wave'' impact models (e.g.
M\'esz\'aros and Rees, 1993a,b), where the fireball sweeps up material
from the surrounding interstellar medium, creating one or several
shocks, much like the propagation of supernova ejecta.  The kinetic
energy of the fireball is then extracted in non-thermal form via
dissipative action of such shocks in the formation of quasi-isotropic
populations of particles through Fermi acceleration.  These particles
efficiently create non-thermal radiation with multiple-component broken
power-law spectra.  M\'esz\'aros and Rees (1993b, see also
M\'esz\'aros, Rees and Papathanassiou 1994) envisaged synchrotron
radiation generally in the optical to X-ray range, and an inverse
Compton ``image'' of the synchrotron continuum in the $\gamma$-ray
range. This scenario yields spectral indices as low as 3/2, which are
quite representative of the EGRET data.  A significant deficiency of
this model is that it does not define a characteristic energy for the
peak of the \teq{\nu}-\teq{F_{\nu}} spectrum that is restricted to the
BATSE energy range.  TeV band observations have the potential to
restrict source parameters such as density \teq{n}, field strength {\bf
B} and bulk Lorentz factor \teq{\Gamma}.

None of the remaining handful of models for super-GeV emission make
detailed spectral predictions nor provide well-constrained flux
estimates.  Katz (1994) suggested that the impact of fireballs with
dense clouds (with \teq{n\gtrsim 10^9}cm) spawned by their progenitors,
perhaps as a wind, could yield gamma-ray emission via \teq{\pi^0}
decay.  Such densities are possible on subparsec scales if the clouds
are sufficiently massive (\teq{\sim M_{\odot}}), and the result would
be a delayed hard gamma-ray signal.  M\'esz\'aros and Rees (1994)
conjecture that the super-GeV emission is a delayed signal from the
impact of the fireball with the ISM while the MeV radiation is
generated by shocks internal to the fireball.  Waxman and Coppi (1996)
proposed that cascading of ultra high energy (\teq{\gtrsim 10^{19}}eV)
cosmic rays off infra-red and cosmic microwave background fields can
generate delayed GeV--TeV emission from cosmological bursts.  This
becomes possible only if the bursts have fields in excess of
\teq{10^5}Gauss, below which Fermi acceleration of cosmic rays to such
high energies is not possible.  While they make no spectral
predictions, the expectations of this model can be assessed from the
calculations that Protheroe and Stanev (1993) performed for such
cascading in the context of active galactic nuclei.  They determined
that flat emission spectra were possible, with cutoffs that were quite
dependent on the source distance.  For this situation, observations by
ACTs may be able to provide distance determinations for bursts.

\section{Spectral Characteristics in the GeV--TeV Band}
While these more-focussed gamma-ray burst models have provided few
spectral predictions, there exist more global calculations that provide
general guides for TeV gamma-ray astronomers of the expectations for
GRB fluxes and spectra.  These results hinge on the transparency or the
opacity of high energy source photons to two-photon pair production
\teq{\gamma\gamma\to e^-e^+}, and have been studied rather extensively
because of their informative, more-or-less model-independent nature.
In the subsequent presentations of this section, bursts will be assumed
to intrinsically emit out to energies of 1 TeV--10 TeV or higher; if
reality should prove otherwise, the relevance of ACT observations to
GRB studies will be very limited.

\vspace{-0.3cm}
\subsection{Opacity to internal pair production}
Attenuation by pair creation in the context of GRBs was first explored
by Schmidt (1978).  He assumed that a typical burst produced
quasi-isotropic radiation, and concluded at the time that the detection
of photons around 1 MeV limited bursts to distances less than a few
kpc, since the optical depth scales as the square of the distance to
the burst.  The EGRET observations of emission above 100 MeV
indicated that Schmidt's analysis needed serious revision, particularly
since BATSE's determination of the spatial isotropy and inhomogeneity
of bursts (e.g.  Meegan et al. 1992) implied that they are either in
an extended halo or at cosmological distances.  Consequently their
intrinsic luminosities, and therefore their optical depths to pair
production for isotropic radiation fields, are much higher than was
previously believed.

In the wake of this apparent conflict, the suggestion (e.g. Fenimore et
al. 1992) that GRB photon angular distributions were highly beamed and
produced by a relativistically moving or expanding plasma emerged.
This hypothesis builds on the property that \teq{\gamma\gamma\to
e^-e^+} has a threshold energy \teq{E_1} that is strongly dependent on
the angle \teq{\Theta} between the photon directions:  \teq{E_1 >
2m_e^2c^4/[1-\cos\Theta] E_2} for target photons of energy \teq{E_2}.
Hence radiation beaming associated with relativistic bulk motion of the
underlying medium can dramatically reduce the optical depth,
\teq{\taupp}, {\it internal} to sources at enormous distances from
earth, suppressing $\gamma$-ray spectral attenuation turnovers, and
blue-shifting them to energies above those detected.  Various
determinations of the bulk Lorentz factor \teq{\Gamma} of the medium
supporting the GRB radiation field have been made in recent years,
mostly concentrating (e.g. Krolik and Pier 1991, Baring 1993, Baring
and Harding 1993) on the simplest case where the angular extent of the
source was of the order of \teq{1/\Gamma}, with an infinite power-law
burst spectrum \teq{n(\erg )=n_{\gamma} \,\erg^{-\alpha }}, where
\teq{\erg} is the photon energy in units of \teq{m_ec^2}.  Under such
assumptions, the pair creation optical depth takes the well-known form
\teq{\taupp (\erg )\propto \erg^{\alpha -1}\Gamma^{-(1+2\alpha )}} for
\teq{\Gamma\gg 1}, so that large Lorentz factors suppress pair creation
very effectively.  Setting the optical depth to unity at the maximum
energy observed by EGRET leads (e.g.  Baring 1993; Harding 1994; Baring
and Harding 1997b) to estimates of \teq{\Gamma\sim 10-30} for galactic
halo sources and \teq{\Gamma\sim 100-1000} for cosmological bursts.
The detailed analysis of pair production transparency for a broad range
of source geometries by Baring and Harding (1997b) revealed that the
optical depth \teq{\taupp (\erg )} was only weakly-dependent on the
opening angle \teq{\Theta} of relativistic expansions when
\teq{\Theta\gtrsim 1/\Gamma}, an effect that is due to restrictions on
the \teq{\gamma\gamma\to e^+e^-} phase space imposed by causality.

\vspace{-0.3cm}
\subsubsection{The effects of sub-MeV spectral curvature on GeV--TeV spectra}
Despite the expedient approximation of using infinite power-law spectra
for most pair production analyses, most bursts detected by BATSE show
significant spectral curvature in the 30 keV--500 keV range (e.g. Band
et al.  1993).  Furthermore, BATSE sees MeV-type (i.e. 500 keV--2 MeV)
spectral curvature with significant frequency in bright bursts,
including EGRET sources (e.g. see Schaefer et al. 1992).  Such
curvature could, in principle, reduce the opacity of potential TeV
emission from these sources, via a depletion of target photons for the
hard gamma-rays.  This important consideration was discussed by Baring
and Harding (1997a), who modelled GRB spectral curvature using broken
power-laws as a first approximation.  They found that the presence of
such curvature generally has minimal influence on the spectra (below 1
TeV) and inferred bulk motions for bursts of cosmological origin.  This
result followed as a consequence of there being a plentiful supply of
target photons (at \teq{E_1\sim m_e^2c^4\,\Gamma^2/E_2}) above the
BATSE range in sources at extragalactic distances.  In contrast, for
galactic halo sources, Baring and Harding (1997a) observed that source
opacity may arise only in a portion of the 1 GeV -- 1 TeV range, with
transparency returning in the super-TeV range, resulting in the
appearance of distinctive, broad absorption troughs.  Such features
would provide a unique identifier for bursts in halo locales.

\begin{figure}[htb]
\vspace{0.2cm}
\centerline{\psfig{file=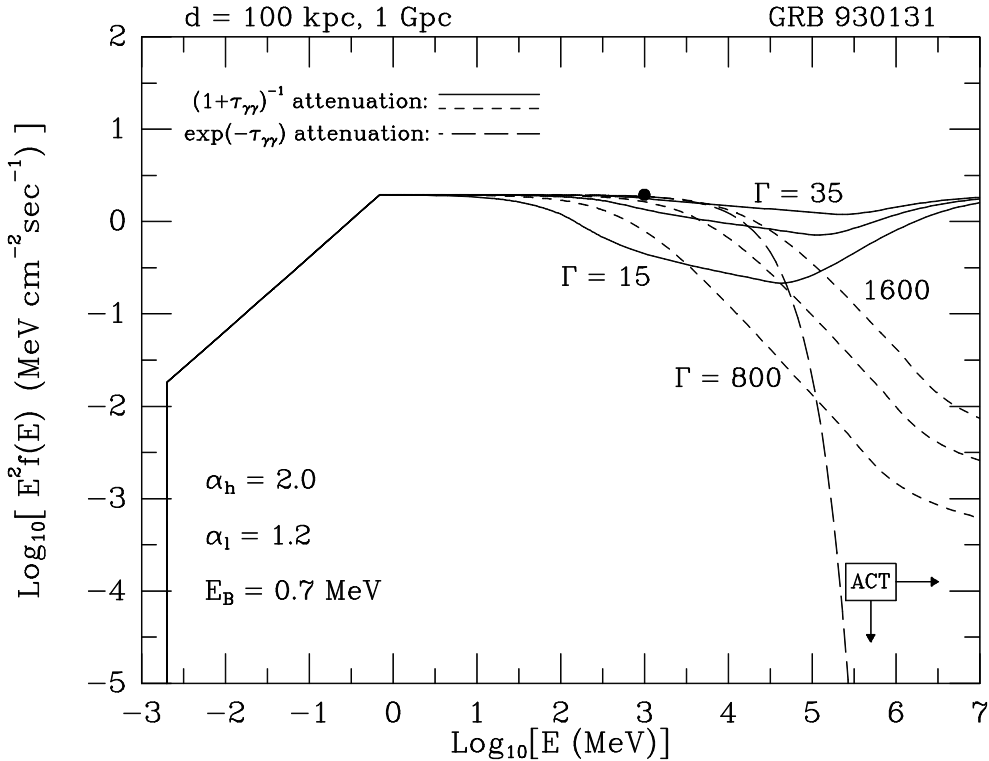,height=10.0cm}}
\vspace{-0.4cm}
\caption{The attenuation, internal to the source, of a broken power-law
spectrum for GRB 930131 at source distances typical of galactic halo
(solid curves, \teq{\Gamma =15,\, 25,\, 25}) and cosmological (short
dashed curves, \teq{\Gamma =800,\, 1200,\, 1600}) origin, and different
bulk Lorentz factors \teq{\Gamma} for the emitting region.  The
spectra, plotted in the \teq{E^2\, f(E)} (i.e. \teq{\nu F_{\nu}})
format, are attenuated by the factor \teq{1/(1+\taupp )} (except for
the \teq{\Gamma=1600}, long dashed line case) for optical depths whose
form is in Baring and Harding (1997a).  The source spectrum was
modelled with a power-law broken at \teq{E_{\rm B} =0.7}MeV, with
spectral indices \teq{\alpha_l=1.2} and \teq{\alpha_h=2.0}.  A low
energy cutoff at 2 keV was used to mimic X-ray paucity.  The filled
circle denotes the highest energy EGRET photon at 1000 MeV (see Sommer
et al. 1994).  The current threshold and sensitivity for ACT
observations of bursts is again indicated by the ``ACT'' box.
}
\vspace{-0.2cm}
\end{figure}

Attenuation of spectra appropriate to the ``Superbowl'' burst GRB
930131 are depicted in Fig.~2 for different \teq{\Gamma}.  Most of the
curves depicted are for attenuation by a factor of \teq{1/(1+\taupp )},
as is appropriate for source photons being distributed in a roughly
spatially-uniform manner (i.e. including ``skin effects'').  The
contrast between spectral shapes for cosmological and galactic halo
bursts is striking.  Absorption troughs appear in the 100 kpc cases and
are quite distinct from the broken power-law structure in cosmological
scenarios.  The spectral indices above 1 GeV in the 1 Gpc examples are
defined uniquely in terms of those at lower energies, a property that
can distinguish this internal absorption from the external photon
absorption described just below.  For both situations, GRB 930131 would
have been easily detectable by ACTs if it had been observed in a slew
search.  An exponential attenuation case (for source distance
\teq{d=1}Gpc) is also illustrated in the figure: similar examples
are given in Baring and Harding (1997a).  This corresponds to
substantial spatial confinement of the target photons, and produces
sharp cutoffs in cosmological scenarios that would render bursts
undetectable by ACTs; halo bursts would still be detectable, however
their absorption troughs would be much more pronounced.  A diversity of
such spectral shapes (e.g. troughs, shelfs and turnovers) might be
anticipated for GRBs.  Clearly future observations and/or upper limits
by ACTs will play a prominent role in constraining burst scenarios and
model parameters such as \teq{\Gamma}.

\vspace{-0.2cm}
\subsection{Absorption due to background fields}
If \teq{\Gamma} happens to be large enough to permit emission out to
TeV energies (it must be \teq{\gtrsim 10^3} for the \teq{d=1}Gpc
example of Figure~2), then spectra from cosmological bursts would
suffer attenuation due to the \it external \rm supply of infra-red (IR)
background (and the cosmic microwave background, CMB) photons (Stecker
and De Jager 1996, Mannheim, Hartmann and Funk 1996).  This issue has
been studied extensively for Mrk 421 and other blazars (e.g. see
Stecker et al., in these proceedings), and the results can be directly
mapped over to gamma-ray bursts.  The absence of attenuation in extant
data can provide upper bounds to the source redshift.  For example,
Stecker and De Jager (1996) concluded that the detection of an 18 GeV
photon from GRB 940217 placed it at a redshift of \teq{z\lesssim 2},
which is not very constraining for cosmological burst populations;
such photons use the CMB as targets.  Clearly, positive detections (or
upper limits) in the TeV band would provide more powerful source
distance diagnostics.

In anticipation of this, Mannheim, Hartmann and Funk (1996) computed
the expected attenuation in GRB spectra for sources at significant
cosmological redshifts \teq{z}, and estimated detection rates for
various TeV and super-TeV gamma-ray experiments.  They contended that
Whipple might expect to see one burst per year, but with its larger
field of view, MILAGRO might detect around ten sources per year.  The
attenuation ``templates'' they produced are exhibited in Figure~3, and
strongly resemble the exponential turnover case in Figure~2:
exponential attenuation is the product of a unique burst distance in
the bath of IR photons.  These templates patently confirm that
telescopes observing only above a few hundred GeV would be unable to
detect bursts with redshifts greater than \teq{\sim 0.5}.  Experiments
such as STACEE and CELESTE may therefore play a crucial part in studies
of bursts if they are at high redshifts, as would be indicated by the
Keck line observations discussed above, and also the time dilation
analysis of Norris et al. (1994) for the BATSE event population.
Note that the model-dependent details of photon/pair cascading (e.g.
see Protheroe and Stanev 1993, for applications to active galactic
nuclei) would probably make no significant qualitative changes to the
shapes of these attenuation templates.  One salient feature of the
sharp spectral turnovers computed by Mannheim, Hartmann and Funk (1996)
is that they depend on \teq{z}, the (at present uncertain) details of
the IR background, and the Hubble constant \teq{H_0}.  The connection of
these quantities may prove valuable in the future.  As absorption
studies coupled with TeV observations of the blazars Mrk 421 and Mrk
501 (see several papers in these proceedings) have made great strides
in constraining the IR background, when our knowledge of this is
sufficiently improved, if optical telescopes like Keck can measure
redshifts \teq{z} to burst counterparts, then constraints on the Hubble
constant can be obtained.  Realistically, it will be some time before
this approach can yield results comparable to other methods for
bounding \teq{H_0}.

\begin{figure}[htb]
\vspace{0.2cm}
\centerline{\psfig{file=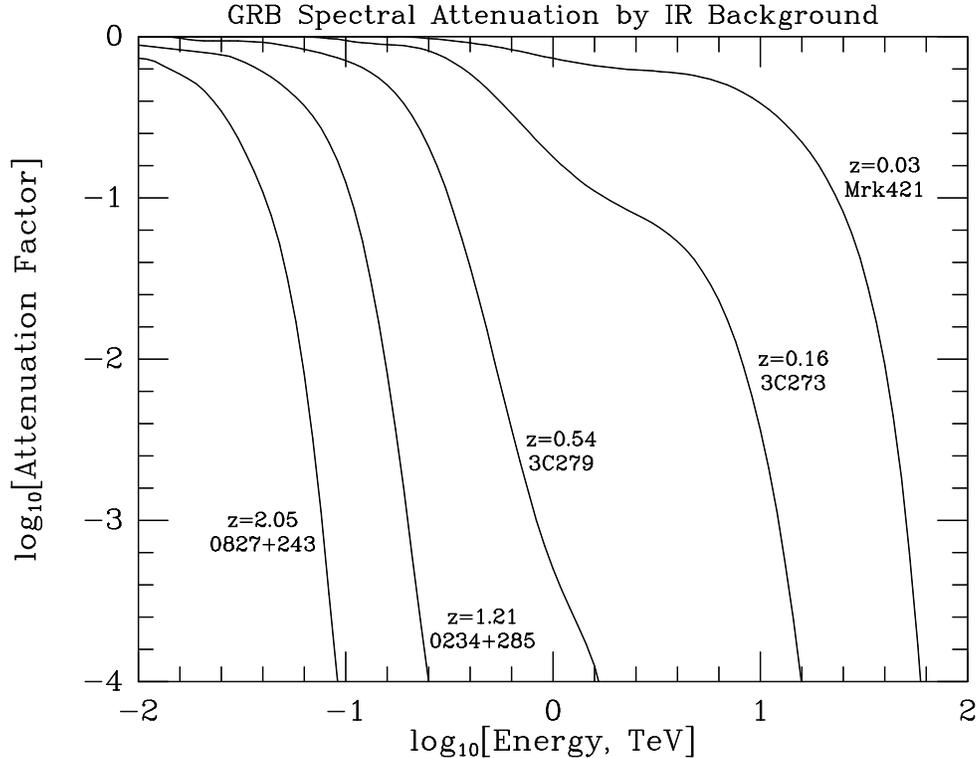,height=10.0cm}}
\vspace{-0.3cm}
\caption{The attenuation factor template, as computed by Mannheim,
Hartmann and Funk (1996, this Figure is an adaptation of their
Figure~3), for absorption of GRB emission by pair production off
external infra-red background photons (they use the IR model of MacMinn
and Primack 1996).  The factors, which are independent of the source
spectral index, are presented for bursts at various redshifts at which
blazars (as labelled) have been detected by EGRET.  The Hubble constant
for these templates was assumed to be \teq{H_0=50}km sec$^{-1}$ Mpc$^{-1}$.
}
\vspace{-0.1cm}
\end{figure}

\section{Conclusion}
The investigation of gamma-ray bursts with the atmospheric \v{C}erenkov
technique is clearly a fledgling field, awaiting the first confirmed
detection.  Yet it is anticipated that, as with their role in studying
other cosmic objects, namely plerions associated with pulsars
(discussed by Harding and De Jager in these proceedings), blazars, and
now shell-type supernova remnants (SN1006: see Tanimori et al., these
proceedings), \v{C}erenkov telescopes will provide ground-breaking
discoveries relating to gamma-ray bursts in due course:  the future for
the TeV gamma-ray astronomy of bursts is bright.  Given the recent Keck
line redshifts from a burst counterpart, it now appears unlikely that
ACTs will play a key role in determining the global distance scale for
gamma-ray bursts.  However, current and future ground-based initiatives
such as Whipple, HEGRA, CANGAROO, CAT, MILAGRO, STACEE and CELESTE, to
name a few, and space missions such as GLAST, will provide key pieces
of information constraining both distances to individual sources, and
burst parameters such as the bulk Lorentz factor \teq{\Gamma}, the
underlying plasma density and magnetic field strength.

Detailed theoretical predictions of the dependence of TeV spectra on
such parameters are currently sparse.  Yet the generic absorption
properties discussed in Section~4 provide strong guidelines for
experimentalists.  It is clear that observations in the 30 GeV--3 TeV
band can discriminate between pair production absorption that is
internal to sources, for which there are strong correlations between
the spectral shapes in different X-ray/gamma-ray bands, and opacity due
to external background fields of radiation.  It is unlikely that
multi-component models of hard gamma-ray emission will replicate the
signatures implied by \teq{\gamma\gamma\to e^+e^-} attenuation.  The
diagnostic capability of existing telescopes to address these issues
may actually be superseded by lower threshold instruments such as
STACEE and CELESTE, which will feature prominently in the near future
given that (i) they probe a key portion of the burst spectral range,
and (ii) their sensitivity (and angular resolution) betters that of
GLAST (their principal competitor) above around 10 GeV.  One very nice
aspect of this problem is that multi-wavelength observations will
maximize the improvement of our understanding of bursts, involving
techniques ranging from the ground-based TeV variety that was the focus
of this conference, space-based hard and soft gamma-ray detectors, and
X-ray telescopes such as that which turned the GRB field around this
year.  It is anticipated that the TeV gamma-ray astronomy community
will be active participants in elucidating the gamma-ray burst mystery
in the future.

\section{Acknowledgements}
I thank my collaborator Alice Harding for numerous insightful
conversations on bursts and for comments helpful to the improvement of
the manuscript, and Brenda Dingus and Jennifer Catelli for many
discussions about EGRET burst data.

\begin{refs}   

\item
Band, D., et al. 1993, \apj\vol{413}{281}
\item
Baring, M.~G. 1993, \apj\vol{418}{391}
\item
Baring, M.~G. and Harding, A.~K. 1993, in Proc. 23rd ICRC (Calgary), \vol{1}{53}
\item
Baring, M.~G. and Harding, A.~K. 1997a, \apjl\vol{481}{L85}
\item
Baring, M.~G. and Harding, A.~K. 1997b, \apj\ in press.
\item
Boyle, P.~J., et al. 1997, in Proc. 25th ICRC (Durban), \vol{3}{61}
\item
Carrigan, B.~J. \& Katz, J.~I. 1992, \apj\vol{399}{100}
\item
Catelli, J.~R. et al. 1996,in \it Gamma-Ray Bursts, \rm eds.  
  Kouveliotou, C., Briggs, M.~F., and Fishman, G.~J. (AIP Conf. Proc. 384, 
  New York) p.~185.
\item
Cavallo, G., and Rees, M.~J. 1978, \mnras\vol{183}{359}
\item
Connaughton, V. et al. 1995, in Proc. 24th ICRC (Rome), \vol{2}{96}
\item
Connaughton, V. et al. 1997, \apj\vol{479}{859}
\item
Costa, E., et al. 1997, \nat\vol{387}{783}
\item
Dazeley, S.~A., et al. 1997, in Proc. 25th ICRC (Durban), \vol{3}{65}
\item
Dingus, B.~L. 1995, \apss \vol{231}{187}
\item
Dingus, B.~L., et al. 1994, in \it Gamma-Ray Bursts, \rm eds.  
  Fishman, G.~J., Hurley, K. and Brainerd, J.~J. (AIP Conf. Proc. 307, 
  New York) p.~22.
\item
Djorgovski, S.~G., et al. 1997, \nat\vol{387}{876}
\item
Eichler, D., et al. 1989, \nat\vol{340}{126}
\item
Fenimore, E. E., et al. 1988, \apjl\vol{335}{L71}
\item 
Fenimore, E.~E., Epstein, R.~I. and Ho, C.: 1992 in \it Gamma-Ray Bursts, \rm 
   eds. Paciesas, W. S. and Fishman, G. J.,
   (AIP Conf. Proc. 265, New York) p. 158.
\item
Fishman, G.~J., et al. 1994, \apjs \vol{92}{229}
\item
Goodman, J. 1986, \apjl\vol{308}{L47}
\item
Groot, P.~J., et al. 1997, IAU Circ. 6584
\item
Guarnieri, A., et al. 1997, \aapl\ in press.
\item
Hanlon, L.~O., et al. 1994, \aap\vol{285}{161}
\item
Harding, A.~K. 1994, in \it Proc. 2nd Compton Symp., \rm ed. Fichtel, C.,
   et al., (AIP Conf. Proc. 304, New York), p.~30.
\item
Hurley, K. 1996, in \it TeV Gamma-Ray Astrophysics, \rm eds. V\"olk, H.~J.
   \& Aharonian, F.~A. (Kluwer, Dordrecht) p.~43.
\item
Hurley, K. et al. 1994, \nat\vol{372}{652}
\item
Katz, J.~I. 1994, \apjl\vol{432}{L27}
\item
Klebesadel, R.~W., Strong, I.~B. and Olson,  R.~A. 1973, \apjl\vol{182}{L85}
\item 
Krolik, J.~H. and Pier, E.~A. 1991, \apj\vol{373}{277}
\item
Kwok, P.~W. et al. 1993, in \it Compton Gamma-Ray Observatory\rm , 
  eds. Friedlander, M., Gehrels, N., and Macomb, D. (AIP Conf. Proc. 280,
  New York) p.~855.
\item
MacMinn, D. \& Primack, J. 1996, \ssr\vol{75}{413}
\item
Mallozzi, R.~S. et al. 1995, \apj\vol{454}{597}
\item 
Mannheim, K., Hartmann, D. and Funk, B. 1996, \apj\vol{467}{532}
\item
Mazets, E.~P. et al. 1981, \apss\vol{80}{3, 85, 119}
\item
Meegan, C., et al. 1992, \nat \vol{355}{143}
\item
M\'esz\'aros, P. \& Rees, M.~J. 1993a, \apj\vol{405}{278}
\item
M\'esz\'aros, P. \& Rees, M.~J. 1993b, \apjl\vol{418}{L59}
\item
M\'esz\'aros, P. \& Rees, M.~J. 1994, \mnras\vol{269}{L41}
\item
M\'esz\'aros, P., Rees, M.~J. \& Papathanassiou, H. 1994, \apj\vol{432}{181}
\item
Metzger, M.~R., et al. 1997, \nat\vol{387}{878}
\item
Nolan, P.~L. et al. 1983, in \it Positron-Electron Pairs in Astrophysics, \rm
   eds. Burns, M.~L., Harding, A.~K., and Ramaty, R., (AIP Conf. Proc. 101,
   New York) p.~59.
\item
Norris, J.~P., et al. 1994, \apj\vol{424}{540}
\item
Paciesas, W.~S., Fishman, G.~J. 1991, eds. Gamma-Ray Bursts (AIP 265, New York)
\item
Paczy\'nski, B. 1986, \apjl\vol{308}{L43}
\item
Padilla, L., et al. 1997, in Proc. 25th ICRC (Durban), \vol{3}{57}
\item
Protheroe, R.~J. \& Stanev, T. 1993, \mnras\vol{264}{191}
\item 
Schaefer, B. E., et al. 1992, \apjl\vol{393}{L51}
\item
Schaefer, B. E., et al. 1994, \apjs\vol{92}{285}
\item 
Schmidt, W. K. H. 1978, \nat\vol{271}{525}
\item
Schneid, E.~J., et al. 1992, \aapl\vol{255}{L13}
\item
Schneid, E.~J., et al. 1996, in \it Gamma-Ray Bursts, \rm eds.  
  Kouveliotou, C., Briggs, M.~F., and Fishman, G.~J. (AIP Conf. Proc. 384, 
  New York) p.~253.
\item
Share, G.~H. et al. 1986, \asr\vol{6(4)}{15}
\item
Shemi, A. \& Piran, T. 1990, \apjl{365}{L55}
\item
Sommer, M., et al. 1994, \apjl \vol{422}{L63}
\item
Stecker, F.~W. \& De Jager, O.~C. 1996, \ssr\vol{75}{401}
\item
van Paradijs, J., et al. 1997, \nat\vol{386}{686}
\item
Waxman, E. \& Coppi, P. 1996, \apjl\vol{464}{L75}
\item
Woosley, S.~E. 1993, \apj\vol{405}{273}
\end{refs}

\end{document}